\documentclass[times,twocolumn]{aastex631}
\usepackage{amsmath}
\usepackage{multirow}
\usepackage{bm}
\usepackage{rotating}
\usepackage{multirow}
\usepackage{nicefrac,xfrac}
\usepackage{mathtools}
\usepackage{scalerel}

\def \wl #1 {\textcolor{red}{#1 }}

\begin{document}

\title{Quantum Statistical Effects on Warm Dark Matter and the Mass Constraint from the Cosmic Structure at Small Scales}

\author[0009-0001-3227-2225]{Zhijian Zhang}
\affiliation{South-Western Institute For Astronomy Research, Yunnan University, Kunming 650500, Yunnan, P. R. China}

\author[0000-0003-2240-7031]{Weikang Lin}
\affiliation{South-Western Institute For Astronomy Research, Yunnan University, Kunming 650500, Yunnan, P. R. China}

\correspondingauthor{Weikang Lin}
\email{weikanglin@ynu.edu.cn}

\begin{abstract}
The suppression of the small-scale matter power spectrum is a distinct feature of Warm Dark Matter (WDM), which permits a constraint on the WDM mass from galaxy surveys. In the thermal relic WDM scenario, quantum statistical effects are not manifest. In a unified framework, we investigate the quantum statistical effects for a fermion case with a degenerate pressure and a boson case with a Bose-Einstein condensation (BEC). Compared to the thermal relic case, the degenerate fermion case only slightly lowers the mass bound, while the boson case with a high initial BEC fraction ($\gtrsim90\%$) significantly lowers it. On the other hand, the BEC fraction drops during the relativistic-to-nonrelativistic transition and completely disappears if the initial fraction is below $\sim64$\%. Given the rising interest in resolving the late-time galaxy-scale problems with boson condensation, a question is posed on how a high initial BEC fraction can be dynamically created so that a condensed DM component remains today.

\end{abstract}

\section{Introduction}\label{sec:introduction}
A wide range of cosmological and astronomical phenomena from galaxy scales
\citep{zwicky1933spectral,smith1936Virgomass,1957-van_de_Hulst,bulletcluster-NASA,Lin:2016vmm,Salucci:2018hqu} to large scales \citep{DES-y1-2018-joint,Hikage-etal-2019,Joudaki-etal-2018,Planck2018-parameter-constraints} point to the existence of dark matter (DM). However, little is known about DM, except that it is around five times the mass of baryons and highly nonrelativistic at the time of recombination (i.e., cold). Here, we assume that dark matter is some unknown particle.% but refer readers to other alternatives such as primordial black holes \citep{Hawking:1971-PBH} and some composite nuggets/objects \citep{Witten:1984rs}.

An important parameter of DM is the particle mass, which is related to almost all aspects of the DM study, such as model building, early-universe production mechanisms, and direct and indirect searches. The cosmic structure, especially at intermediate to small scales, has been a powerful probe of DM properties, especially the ``warmness'' of DM. The early-time thermal motion of DM smooths out the structures and prevents gravitational growth at small scales. In the standard thermally produced relic scenario, WDM became a free stream when it was still relativistic. Matching the current DM energy density requires that lighter DM be warmer. So, the nondetection of the suppression of the matter power spectrum places a lower bound on the WDM mass \citep{Colombi-Scott-Widrow-1996}, which is currently of a few keV from the Lyman-$\alpha$ forest \citep{Irsic:2023equ,Villasenor:2022aiy}. While we focus on the Lyman-$\alpha$ constraint, we remind readers that the suppression of the matter power spectrum, and thus the WDM mass, can also be constrained by other observations. These include the abundance of small halos and subhalos as probed by strong lensing \citep{Li:2016mnras}, subhalo-star stream interactions \citep{Erkal:2015mmnra_I,Erkal:2015mmnra_II}, satellite galaxy abundance \citep{Nadler:2021PhRvL,2020-Newton-etal}, the high-redshift luminosity function \citep{Rudakovskyi:2021jyf}, and, potentially, the global and anisotropic 21cm signals \citep{Boyarsky-etal-2019-WDM-21cm,Chatterjee:2019mnras}. 

The above constraint on the DM mass is not absolute but depends on the assumption of the DM production mechanism \citep{BhupalDev:2013oiy}. In particular, while a fermion WDM is assumed in the usual thermal relic scenario, quantum statistical effects only play a minor role. When the WDM number density is larger than that in the thermal relic scenario, quantum statistical effects begin to be significant. For the fermion case, a degenerate pressure is added on top of the thermal pressure \citep{Bar:2021jff,Carena:2021bqm}. For the boson case, a Bose-Einstein condensation (BEC) component appears \citep{Madsen:1992am} and can affect the cosmic background evolution and structure formation \citep{Fukuyama:2006gp,Harko:2011zt,Chavanis:2012aap}. 

Given the importance of the WDM mass constraint and the growing interest in quantum effects on galaxy and cosmological scales, in this work, we investigate how quantum statistical effects impact the mass constraint from the Lyman-$\alpha$ forest. From the view of conservation of specific entropy, we analyze the evolution of quantum statistical effects during the relativistic-to-nonrelativistic transition (RNRT) and their impacts on RNRT and the matter power spectrum. 
%This is important for the early-time structure growth but has not been systematically studied.
In particular, compared to previous works on cosmological DM condensation \citep{Fukuyama:2006gp,Harko:2011zt}, we focus on how the BEC fraction evolves during RNRT and the effects on the WDM mass constraint from cosmic structure. 
The scenario considered in this work serves as a minimum extension to the standard thermal relic WDM that considers both a degenerate fermion case and a boson case with a BEC component. We dub it quantum WDM (qWDM).

We adopt the units $G=c=k_{\rm B}=1$.

\section{Analysis}\label{sec:analysis}
We consider a single-species ideal-gas DM that has an equilibrium phase-space distribution,
\begin{equation}\label{eq:HiDM-distribution}
    f(p;T,\mu) = \frac{g}{2\pi^2\hbar^3}\frac{p^2}{\exp\big[(\sqrt{p^2+m^2}-\mu)/T\big]\pm1}\,,
\end{equation}
with a mass $m$, intrinsic degree of freedom $g$, temperature $T$ and chemical potential $\mu$. In the denominator, $+1$ is for fermion and $-1$ for boson. Note that this DM temperature $T$ is different from the standard-sector temperature. 
In this work, we do not assume any early thermal contact between DM and the standard model, and thus DM remains hidden.\footnote{It is more suitable to call our scenario the ``Hidden Dark Matter'' \citep{Chen-Tye-2006}, which is more generalized. We adopted the term WDM to better highlight the quantum statistical effects compared to the thermal relic scenario.} Since we assume that DM has an equilibrium phase-space distribution instead of being a free stream, it requires some DM-DM interaction to keep DM in equilibrium at least in the early time \citep{Egana-Ugrinovic:2021gnu}. We take $g=2$ for the fermion case and $g=1$ for the boson case, but the conclusions obtained in this work can be readily generalized to cases with a higher $g$. 

We define the following two dimensionless variables:
\begin{align}
    &x\equiv \frac{m}{T}\,, \\
    &\Delta\equiv\frac{\mu-m}{T}\,.
\end{align}
Then $x$ is the mass-to-temperature ratio, and $\Delta$ is the effective chemical potential-to-temperature ratio. We call $\mu-m$ the effective chemical potential for the following reason. In the relativistic limit, the mass $m$ can be ignored. In the nonrelativistic limit, the distribution Eq.\,\eqref{eq:HiDM-distribution} reduces to the nonrelativistic form, from which we can identify $\mu-m$ as the nonrelativistic chemical potential.

From Eq.\,\eqref{eq:HiDM-distribution}, we calculate the physical particle number density $n$, energy density $\rho$, and the pressure $P$, as well as the entropy density $s$ for given temperature and chemical potential; see Appendix \ref{sec:method-rel-nonrel-limits}. Some important results for $n$, $\rho$, $P$, and $s/n$ in the relativistic and nonrelativistic limits are summarized in Table \ref{tab:rel_and_nonrel_limits}. Those limits are useful for obtaining the relations between the initial and final values of dynamical variables and verifying our numerical solutions of the qWDM background evolution.

The parameter $\Delta$ is an important parameter in determining the degree of quantum statistical effects. The system is called classical when $\Delta\rightarrow-\infty$ and Eq.\,\eqref{eq:HiDM-distribution} reduces to the Boltzmann-Maxwell distribution. For the fermion case, the degenerate pressure begins to play a role when $\Delta>0$ and the system is highly degenerate when $\Delta\rightarrow\infty$. For the boson case, a Bose-Einstein condensation (BEC) can take place when $\Delta=0$. %\citet{Lin:2023fao} showed that $\Delta$ drops from one constant in the relativistic limit to another constant in the nonrelativisticlimit. The initial conditions are then the value of 
In that case, we define the fraction of the BEC component as
\begin{equation}\label{eq:BEC-fraction}
    r\equiv \frac{n_{\textsc{bec}}}{n_{\rm tot}}\,,
\end{equation}
where $n_{\textsc{bec}}$ is the particle number density of qWDM in the BEC state and $n_{\rm tot}=n+n_{\textsc{bec}}$ is the total qWDM number density. We assume that there is no internal energy state, and the BEC component of DM condenses into the zero-momentum state so that the total energy density is 
\begin{equation}\label{eq:total-energy-density-BEC}
    \rho_{\rm tot}=\rho + mn_{\textsc{bec}}\,.
\end{equation}
Note that $n_{\rm tot}=n$ and $\rho_{\rm tot}=\rho$ for the fermion case.

The background (homogeneous level) evolution of the qWDM thermodynamical variables, e.g., $\rho(a)$ and $P(a)$, is solved assuming the conservation of comoving particle number and the conservation of specific entropy, that is,\footnote{Note that, with the particle number conversation, the qWDM evolution obtained based on the conservation of specific entropy is equivalent to that obtained based on the energy conservation. However, using the conservation of specific entropy allows us to see clearly how the quantum statistical effects evolve, as we will discuss. }
\begin{align}
    n_{\rm tot}a^3 &= {\rm constant}\,,\\
    s/n_{\rm tot} &= {\rm constant}\,,
\end{align}
where $a$ is the scale factor. See Appendix \ref{sec:method-rel-nonrel-limits} for the details of the numerical calculation of the background evolution.

\subsection{The DM energy density today}
For the fermion case, the initial temperature and chemical potential determine the thermal WDM number density. For the boson case with a BEC component, the initial chemical potential vanishes, but the initial fraction of the BEC component $r_{\rm i}$ is a free parameter. For both cases, the total DM energy density today is the product of the total number density and the mass. It can be shown that the DM energy density fraction today is given by
\begin{equation}\label{eq:DM-fraction-alpha}
    \Omega_{\rm dm} h^2 = \frac{g}{2}\beta(\Delta_{\rm i}, r_{\rm i})\frac{m}{94\,{\rm eV}}\frac{\alpha^3}{4/11}\,,
\end{equation}
where
\begin{equation}\label{eq:beta-factor}
    \beta(\Delta_{\rm i}, r_{\rm i})\equiv\frac{J_3^\mp(\Delta_{\rm i})}{(1-r_{\rm i})J_3^-(0)}\,,
\end{equation}
$\alpha\equiv\frac{T_0}{T_\gamma^0}$ with $T_0\equiv\lim_{a\rightarrow0}aT$ and $T_\gamma^0$ is the photon temperature today and
\begin{equation}\label{eq:J_s-equation}
    J_s^\mp(\Delta)\equiv\int_0^\infty\frac{z^{s-1}{\rm d}z}{\exp(z-\Delta)\pm1}\,.
\end{equation} 
%$J_s^\mp(\Delta)\equiv\int_0^\infty\frac{z^{s-1}{\rm d}z}{\exp(z-\Delta)\pm1}$}. %Please see Table \ref{tab:rel_and_nonrel_limits} for the definition of the $J$'s functions. 
Now, the superscript ``$-$" is for fermion and ``$+$" for boson. Note that, while we matched the form of Eq.\,\eqref{eq:DM-fraction-alpha} to that of Eq.\,(6) in \citet{Colombi-Scott-Widrow-1996}, the $\beta$ here is determined by the initial effective chemical potential-to-temperature ratio and the initial fraction of BEC component instead of an arbitrary factor in the momentum distribution. Unlike in the free-streaming case, the $f(p)$ in our scenario follows an explicit nonrelativistic equilibrium form. Consequently, $T_0$ differs from the current temperature of qWDM. For the same initial temperature, both a higher degree of degeneracy for the fermion case and a higher BEC fraction for the boson case give a higher $\beta$ factor and a higher DM density today. However, $\alpha$ is not directly related to an observable, and we shall replace it with the RNRT scale factor, which predominantly determines the suppression scale of the matter power spectrum.

\subsection{The relativistic-to-nonrelativistic transition}\label{sec:transition}
After we solve for the background evolution, the adiabatic sound speed is calculated by
\begin{equation}\label{eq:adiabatic-sound-speed}
    c_{\rm s}=\sqrt{\frac{{\rm d}P/{\rm d}a}{{\rm d}\rho_{\rm tot}/{\rm d}a}}\,.
\end{equation}
%and implemented into the linear perturbation and the public Boltzmann code \textsc{camb} \citep{Lewis-Challinor-Lasenby-2000} to study the effects on the matter power spectrum. We investigate the effects of a different initial degree of degeneracy for the fermion case and a different initial fraction of the BEC component for the boson case. 
In general, qWDM starts with a relativistic phase where $c_{\rm s}$ equals $1/\sqrt{3}$ and then transitions to a nonrelativistic phase where it drops as $1/a$. We define the transition scale factor $a_{\rm nr}$ by the following asymptotic behavior of $c_{\rm s}$ in the nonrelativistic limit:
\begin{align}\label{eq:transition-scale-cs}
    c_{\rm s}\xrightarrow[]{\rm nonrelativistic}\frac{1}{\sqrt{3}}\frac{a_{\rm nr}}{a}\,.
\end{align}
That is, $a_{\rm nr}$ is the scale factor at which the nonrelativistic sound speed would increase to $1/\sqrt{3}$ if it kept increasing inverse-linearly with $a$ when going back in time. %The velocity dispersion of WDM in the early universe smooths out and suppresses the cosmic structure at scales smaller than a suppression scale $\ell_{\rm s}$. The dominant factor determining $\ell_{\rm s}$ is $a_{\rm nr}$. The larger $a_{\rm nr}$, the larger $\ell_{\rm s}$. 

\begin{figure}[tp]
    \centering
    \includegraphics[width=\linewidth]{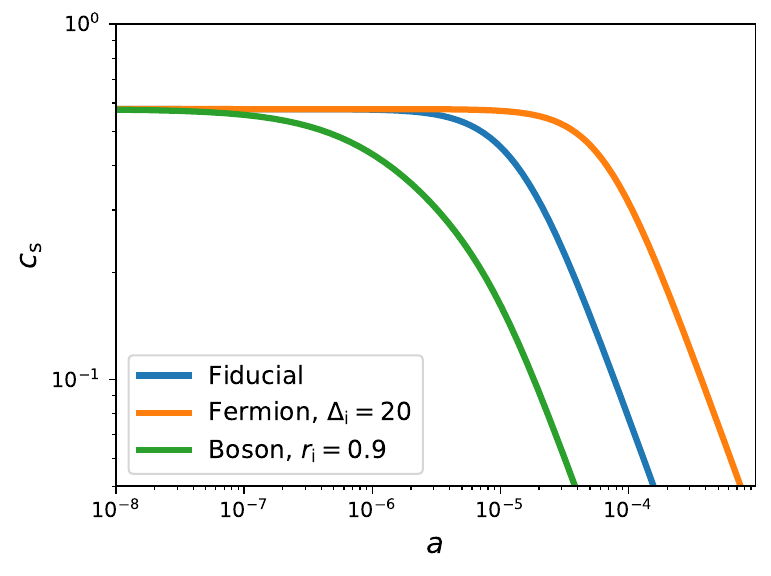}
    \caption{Effects of quantum statistics on the adiabatic sound speed. The degenerate pressure postpones the RNRT for the fermion case, while a BEC component for the boson case expedites it. The same initial value of $x/a$ is applied.}
    \label{fig:sound-speed-comparison}
\end{figure}

Different quantum statistics have distinct effects on $a_{\rm nr}$. The fiducial case corresponding to (but somewhat different from) the thermal relic WDM scenario is a fermion case with a vanishing initial chemical potential and $g=2$. This case has only a small difference from the classical case on the value of $a_{\rm nr}$ for the same initial conditions \citep{Lin:2023fao}. We show the evolution of $c_{\rm s}$ for the classical case with the orange curve in Figure \ref{fig:sound-speed-comparison}. 

For the fermion case with a positive initial $\Delta$ (denoted as $\Delta_{\rm i}$), the degenerate pressure postpones RNRT with a larger $a_{\rm nr}$ compared to the classical case. The larger $\Delta_{\rm i}$, the larger $a_{\rm nr}$.  The reason is the following. In the presence of the degenerate pressure, the condition $m\gg T$ (or $x\gg 1$) is no longer sufficient to define the nonrelativistic phase, as it would be in the classical case. If the degenerate pressure is high enough, a large fraction of DM particles can still occupy relativistic momentum states. Mathematically, only when $x\gg\Delta$ will the momenta of most particles be smaller than the mass and Eq.\,\eqref{eq:HiDM-distribution} be approximated by a nonrelativistic form. Therefore, the nonrelativistic limit requires both $x\gg 1$ and $x\gg \Delta$; that is, in addition, the particle mass needs to be much larger than the effective chemical potential. During the relativistic phase, $x$ increases linearly with $a$, but $\Delta$ remains a constant. Only when $x$ increases to be larger than $\Delta$ will DM enter the nonrelativistic phase. As a result, the higher the initial degree of degeneracy (i.e., the larger $\Delta_{\rm i}$), the later the transition. Such an effect is shown by the blue curve in Figure \ref{fig:sound-speed-comparison}.

For the boson case with a fraction of the BEC component, the opposite effect takes place, and the transition is advanced compared to the classical case. This is because now $x\ll1$ is not the criteria for the relativistic phase. When $x\ll1$ with the presence of a BEC component, the total energy density is,
\begin{equation}\label{eq:energy-density-BEC}
    \rho_{\rm tot}=\rho+mn_{\textsc{bec}}=\rho\left(1+\frac{rx}{1-r}\frac{J_3^+(0)}{J_4^+(0)}\right)\,,
\end{equation}
where both $J_3^+(0)$ and $J_4^+(0)$ are of the order of unity. Meanwhile, $P=\rho/3$ is still satisfied for $x\ll1$. From Eq.\,\eqref{eq:energy-density-BEC}, we can see that when $x\sim(1-r)/r$, the term $mn_{\textsc{bec}}$ is comparable to $\rho$ and the pressure. This estimate breaks down at small $r$ values where higher-order terms in $x$ are needed. Numerical analyses suggest that such an estimate is valid for $r\sim1$ and that $x\ll(1-r)$ is a better criterion for the relativistic phase for all $r>0$. Then, once $x$ increases to $(1-r)$, qWDM already begins to enter the nonrelativistic phase. Therefore, the larger the $r_{\rm i}$, the earlier the RNRT.

%This is mainly because the temperature drops faster during the transition compared to the classical case. For the same value of $\Delta$ (we have $\Delta=0$ when there is a BEC component), the thermal specific entropy decrease during RNRT. This requires the BEC fraction drops to keep the specific entropy  constant; see Sec.\,\ref{sec:transition}. Effectively, this can be seen as heat transferring from the thermal component to the BEC component. As a result, the temperature drops faster than that in the non-BEC case which leads to an advanced transition.

\begin{figure}[tp]
    \centering
    \includegraphics[width=\linewidth]{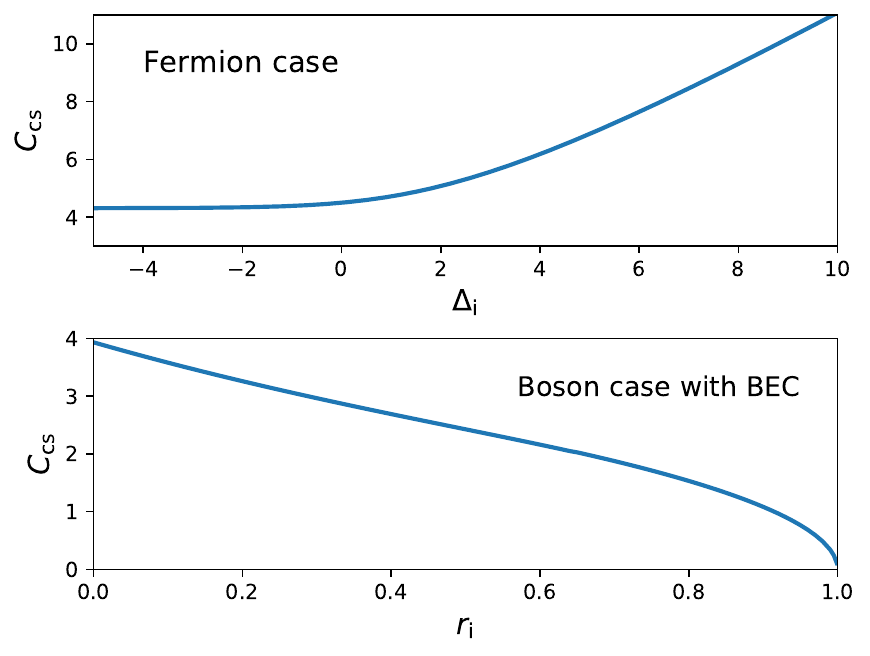}
    \caption{The effect of quantum statistics on the coefficient $C_{\rm cs}$ defined in Eq.\,\eqref{eq:a_hat-related-to-initial}. The larger the $C_{\rm cs}$, the later the RNRT for a given initial mass-to-temperature ratio.}
    \label{fig:Ccs}
\end{figure}

For all cases, the transition scale is proportional to the initial warmness, that is,
\begin{equation}\label{eq:a_hat-related-to-initial}
    a_{\rm nr}=C_{\rm cs}(\Delta_{\rm i}, r_{\rm i})\frac{T_0}{m}\,.
\end{equation}
The coefficient $C_{\rm cs}$ depends on the initial degree of degeneracy for the fermion case and the initial BEC fraction for the boson case. It can be derived analytically. This is done first by substituting the nonrelativistic forms of $\rho$ and $P$ (see Table \ref{tab:rel_and_nonrel_limits}) into Eq.\,\eqref{eq:adiabatic-sound-speed} to relate $c_{\rm s}$ and $x$ in the nonrelativistic phase. Then, the asymptotic evolution of $x$ at the nonrelativistic phase can be derived by matching the initial and the final comoving particle density. As a result, it can be shown that
\begin{equation}\label{eq:Ccs-equation}
     C_{\rm cs}=\left(\frac{5}{3}\right)^{\nicefrac{1}{2}}\frac{(1-r_{\rm f})^{\nicefrac{5}{6}}}{(1-r_{\rm i})^{\nicefrac{1}{3}}}\frac{\big(J_3^\mp(\Delta_{\rm i})\big)^{\nicefrac{1}{3}}\big(J_{\nicefrac{5}{2}}^\mp(\Delta_{\rm f})\big)^{\nicefrac{1}{2}}}{\big(J_{\nicefrac{3}{2}}^\mp(\Delta_{\rm f})\big)^{\nicefrac{5}{6}}}\,.
\end{equation}
The larger the $C_{\rm cs}$, the later the transition. We show $C_{\rm cs}$ as function of $\Delta_{\rm i}$ and $r_{\rm i}$ in Figure \ref{fig:Ccs}.
By combining Eqs.\,\eqref{eq:DM-fraction-alpha} and \eqref{eq:a_hat-related-to-initial}, we relate the DM energy fraction to the transition scale factor, which reads
\begin{equation}\label{eq:DM-fraction-a-hat-cs}
    \Omega_{\rm dm} h^2 = 0.12\times\frac{g}{2}\,\gamma(\Delta_{\rm i}, r_{\rm i})\left(\frac{m}{m_0}\right)^4\left(\frac{a_{\rm nr}}{\hat{a}_{\rm nr}}\right)^3\,,
\end{equation}
where $m_0=1.1\,{\rm keV}$, $\hat{a}_{\rm nr}=1.5\times10^{-7}$ and
\begin{equation}\label{eq:gamma-factor}
    \begin{split}
        \gamma(\Delta_{\rm i}, r_{\rm i})&=\beta(\Delta_{\rm i},r_{\rm i})\Big(\frac{C_{\rm cs}^0}{C_{\rm cs}}\Big)^3\\
        &=\left(\frac{J_{\nicefrac{3}{2}}^\mp(\Delta_{\rm f})}{(1-r_{\rm i})J_{\nicefrac{3}{2}}^-(\Delta_{\rm f}^0)}\right)^{\nicefrac{5}{2}}\left(\frac{J_{\nicefrac{5}{2}}^-(\Delta_{\rm f}^0)}{J_{\nicefrac{5}{2}}^\mp(\Delta_{\rm f})}\right)^{\nicefrac{3}{2}}\,,
    \end{split}
\end{equation}
where $C_{\rm cs}^0$ and $\Delta_{\rm f}^0$ are the values of $C_{\rm cs}$ and $\Delta_{\rm f}$ in the fiducial case (i.e., a fermion case with $\Delta_{\rm i}=0$). The value of $\hat{a}_{\rm nr}$ is motivated by the discussion of the suppression scale in the next section.

\subsection{The suppression scale of the matter power spectrum}\label{sec:suppression-scale}
\begin{figure}
    \centering
    \includegraphics[width=0.995\linewidth]{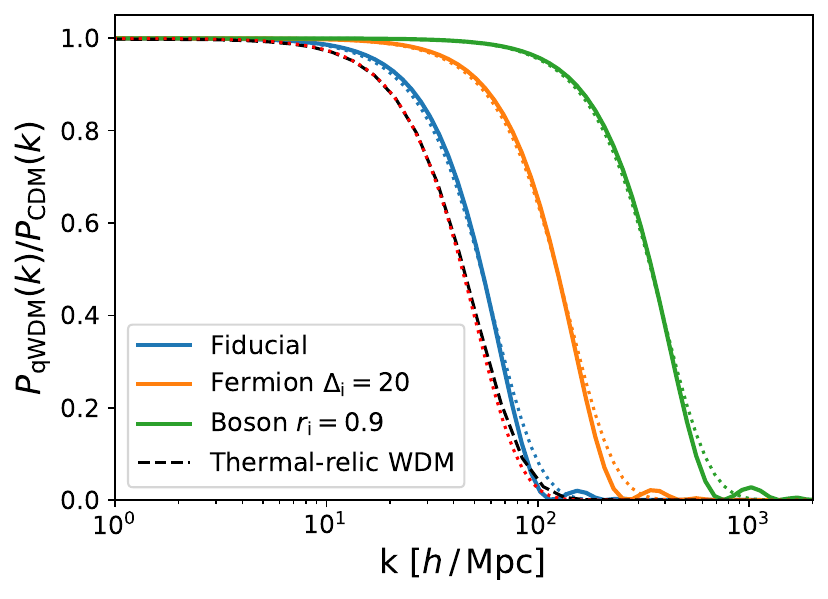}
    \caption{The ratio of matter power spectrum between some selected qWDM cases and the CDM case (solid curves). We set $m=5$\,keV for all cases. The black-dashed curve corresponds to the standard thermal relic WDM with the same mass. The dotted curves are approximations obtained with Eqs.\,\eqref{eq:suppression-scale-numeric}, \eqref{eq:Viel-suppression} and \eqref{eq:alpha-ls}.  The red dotted curve is obtained by the same procedure but with the DM mass scaled by a factor of $\frac{1}{0.84}$; see Eq.\,\eqref{eq:mass-mapping}.} 
    \label{fig:transfer-function}
\end{figure}

To obtain the theoretical prediction of the (comoving) suppression scale of the matter power spectrum, we first estimate it by the qWDM sound horizon scale,
\begin{align}\label{eq:sound_horizon}
    \ell_{\rm s}^{\rm est} &= \int c_{\rm s} {\rm d}\eta=\int\frac{c_{\rm s}}{a^2H}{\rm d}a  \nonumber\\
    &\simeq\frac{1}{\sqrt{3}}\left(\int_0^{a_{\rm nr}}\frac{{\rm d}a}{a^2H}+\int_{a_{\rm nr}}^{a_{\rm eq}}\frac{a_{\rm nr}{\rm d}a}{a^3H}+\int_{a_{\rm eq}}^{1}\frac{a_{\rm nr}{\rm d}a}{a^3H}\right)  \nonumber \\
    &\simeq234\,{\rm Mpc}\times \frac{a_{\rm nr}}{a_{\rm eq}}\Big( 1 - \frac{1}{3}\ln\frac{a_{\rm nr}}{a_{\rm eq}}\Big)\,.
\end{align}
We then follow the procedure in \citet{Lin:2023fao} to implement the adiabatic sound speed in the linear perturbation and the public Boltzmann code \textsc{camb} \citep{Lewis-Challinor-Lasenby-2000} and numerically calculate the matter power spectrum today. We define the suppression scale $\ell_{\rm s}$ as the scale where the matter power spectrum in the qWDM case is half of that in the CDM case, that is,
\begin{equation}\label{eq:suppression-scale-camb}
    \frac{P_{\rm qWDM}(\ell_{\rm s})}{P_{\rm CDM}(\ell_{\rm s})}=\frac{1}{2}\,.
\end{equation}
The fact that the estimation $\ell_{\rm s}^{\rm est}$ is only $\sim2$ times smaller than the numerical $\ell_{\rm s}$ motivates us to parameterize $\ell_{\rm s}$ as
\begin{equation}\label{eq:suppression-scale-numeric}
    \ell_{\rm s}=1\,{\rm Mpc}\times\frac{a_{\rm nr}}{A}\big(1-B\ln(\frac{a_{\rm nr}}{A})\big)\,.
\end{equation}
We find that $A=\hat{a}_{\rm nr}=1.5\times10^{-7} $ and $B=0.145$ make Eq.\,\eqref{eq:suppression-scale-numeric} fit the numerical results well.\footnote{Besides $a_{\rm nr}$, which predominantly determines the suppression scale, different quantum statistics also make the transition width somewhat different; see \citet{Lin:2023fao}. While we ignored such a small difference, including it does not qualitatively change our conclusions. } For a given qWDM mass, $a_{\rm nr}$ can be inferred from Eq.\,\eqref{eq:DM-fraction-a-hat-cs}, which is put in Eq.\,\eqref{eq:suppression-scale-numeric} to give $\ell_{\rm s}$. Despite the different physical settings, the $\ell_{\rm s}$ inferred for a given mass with $\Delta_{\rm i}=0$ and $g=2$ is only a factor of $\sim1.5$ different compared to that in the free-streaming thermal relic WDM scenario provided in other works, such as \citet{Bode-2001, Hansen:2001zv,Viel:2013fqw}. Figure \ref{fig:transfer-function} shows the ratios of the matter power spectrum between the qWDM and CDM cases, both with a DM mass of  $m = 5$\,keV. The cutoff shape for different qWDM scenarios is similar to that of the standard thermal relic WDM. At the smallest scales, the matter power spectrum for qWDM exhibits oscillations. While this feature is not currently relevant for observations, it may become significant as smaller scales are explored in the future. 

To make the connection between the qWDM mass and $\ell_{\rm s}$ clearer, we combine Eqs.\,\eqref{eq:DM-fraction-a-hat-cs} and \eqref{eq:suppression-scale-numeric} to obtain
\begin{equation}\label{eq:DM-fraction-suppression-scale}
    \Omega_{\rm dm} h^2 =0.12\times\frac{g}{2}\gamma(\Delta_{\rm i}, r_{\rm i})\left(\frac{m}{m_0}\right)^4\left[K_{\rm inv}\Big(\frac{\ell_{\rm s}}{1\,{\rm Mpc}}\Big)\right]^3\,,
\end{equation}
where $K_{\rm inv}$ is the inverse function of 
\begin{equation}\label{eq:K-function}
    K(y)=y\,(1-0.145\ln y)\,.     
\end{equation}
Importantly, since the Lyman-$\alpha$ forest constrains $\ell_{\rm s}$, from Eq.\,\eqref{eq:DM-fraction-suppression-scale} we can see that the quantum statistical effects on the constraint of the WDM mass are reflected in $\gamma$. Then, from Eq.\,\eqref{eq:gamma-factor}, there are the two dominant factors that affect $\gamma$ and hence the mass constraint: (1) $\beta$, which is the ratio of the number density between the qWDM case and the thermal relic WDM case, and (2) $C_{\rm cs}$, which determines the time of RNRT.

\section{The mass constraints}\label{sec:mass-constraints}
With the above discussions, it is now ready to see the effects of quantum statistics on the WDM mass constraint from the Lyman-$\alpha$ forest. So far, the suppression on the small-scale matter power spectrum has not been observed, and it thus puts a lower limit on the WDM mass. To apply the current Lyman-$\alpha$ forest constraints on the suppression scale, we first note that the ``transfer function'' of qWDM can be parameterized similarly to that of thermal relic WDM by \citep{Bode-2001}
\begin{equation}\label{eq:Viel-suppression}
    \left(\frac{P_{\rm WDM}}{P_{\rm CDM}}\right)^{\nicefrac{1}{2}}\simeq\left(1+(\alpha k)^{2\mu}\right)^{\nicefrac{-5}{\mu}}\,,
\end{equation} 
with $\mu=1.12$ \citep{Viel-etal-2005}. With our definition of the suppression scale, we can relate $\alpha$ to $\ell_{\rm s}$ by
\begin{equation} \label{eq:alpha-ls}
    \alpha = \frac{\ell_{\rm s}}{2\pi}\big(2^{\nicefrac{\mu}{10}}-1\big)^{\nicefrac{1}{2\mu}}\,.
\end{equation} With $\ell_{\rm s}$ solved in Eq.\,\eqref{eq:DM-fraction-suppression-scale}, the parameterized transfer functions (squared) for different cases are represented by the dotted curves in Figure \ref{fig:transfer-function}. The parameterized transfer function matches the overall suppression effect of the numerical result very well. 

\begin{figure}[tp]
    \centering
    \includegraphics[width=\linewidth]{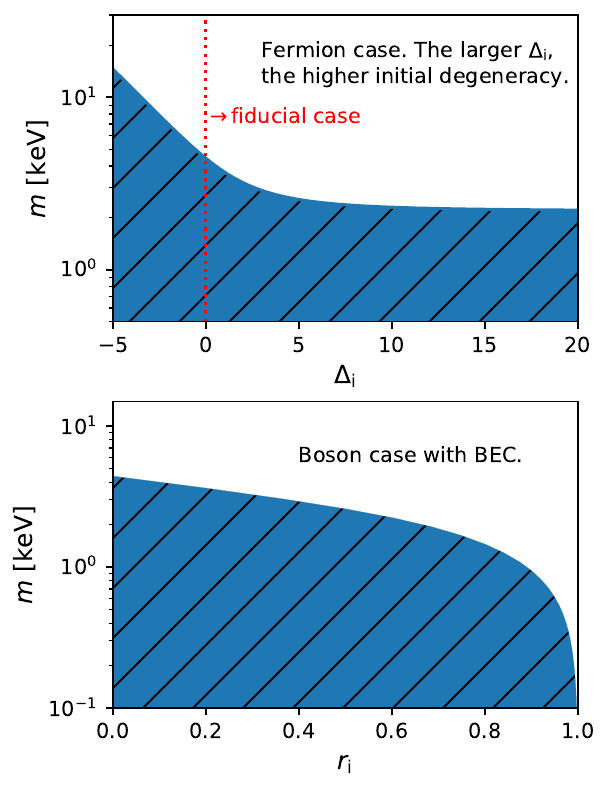}
    \caption{Estimated DM mass constraints from the Ly$\alpha$ forest. Hatched parameter regions are excluded. Top: the fermion case. The mass lower bound is stronger at the classical limit but becomes constant at the highly degenerate limit. The red-dotted vertical line corresponds to the fiducial case.  Bottom: the boson case with a fraction of the BEC component. The mass lower bound is significantly lowered at a high BEC fraction. For an initial BEC fraction close to $100\%$, the mass lower bound can be orders of magnitude below than what is presented here.} 
    \label{fig:mass-constraints}
\end{figure}

Recently, a strong constraint of $\ell_{\rm s}$ was given in \citet{Irsic:2023equ} that the matter power spectrum cannot drop more than $5\%$ at the wavenumber of $14.35\,h$Mpc$^{-1}$.
This corresponds to $\ell_{\rm s}<0.135\,h^{-1}$Mpc by using Eq.\,\eqref{eq:Viel-suppression}. We take $h=0.674$ and $\Omega_{\rm dm}h^2=0.12$ \citep{Planck2018-parameter-constraints}. According to Eq.\,\eqref{eq:DM-fraction-suppression-scale}, the upper bound of $\ell_{\rm s}$ is then translated into a lower bound of the WDM mass that depends on the degree of quantum statistics. In Figure \ref{fig:mass-constraints}, we show the constraints of the qWDM mass as a function of $\Delta_{\rm i}$ for the fermion case and $r_{\rm i}$ for the boson case.

For the fiducial case, the mass is constrained to be $m>4.6 $\,keV. This constraint is somewhat different from $m>5.7$\,keV given in \citet{Irsic:2023equ}, as we assume WDM maintains an equilibrium phase-space distribution rather than being a free stream; also see \citet{Egana-Ugrinovic:2021gnu,Garani:2022yzj}. Additionally, the precision of the usual mass-to-suppression scale parameterization at the large-mass region contributes to this difference \citep{Vogel:2022odl}. The classical case with $\Delta_{\rm i}\ll-1$ makes the mass lower bound stronger (i.e., higher). This is because the DM particle number is smaller in the classical case than in the fiducial case, and then the $\beta$ factor is smaller. Meanwhile, the $C_{\rm cs}$ factor is similar as long as $\Delta_{\rm i}<0$. These result in a smaller $\gamma$ factor in Eq.\,\eqref{eq:DM-fraction-suppression-scale} and a higher mass bound in the classical case compared to the fiducial case. % The DM energy density today is then smaller with a smaller $\gamma$ factor in Eq.\,\eqref{eq:DM-fraction-suppression-scale}. 
  
For the fermion case with the presence of degenerate pressure, the mass constraint is only slightly lower than the fiducial case. Two effects occur that almost cancel each other out and leave the mass constraint almost unchanged. First, DM has a higher number density compared to the fiducial case for a given initial warmness,  which raises the $\beta$ factor in Eq.\,\eqref{eq:DM-fraction-alpha} and tends to lower the mass constraint. But, as discussed in Sec.\,\ref{sec:transition}, RNRT is postponed due to the degenerate pressure so that $C_{\rm cs}$ is also larger. Both $\beta$ and $C_{\rm cs}^3$ have the same asymptotic dependence on $\Delta_{\rm i}$, i.e., $\propto\Delta^3_{\rm i}$ when $\Delta_{\rm i}\gg1$. As a result, the $\gamma$ factor becomes a constant, and the lower bound of the qWDM mass becomes independent of $\Delta_{\rm i}$ at the highly degenerate limit. Despite the different analyses here, the conclusion that fermionic degenerate pressure only slightly lowers the WDM mass constraint is consistent with other works; see \citep{Bar:2021jff,Carena:2021bqm}.

For the boson case with a BEC component, the mass constraint can be significantly lowered with a high BEC fraction. Similar to the fermion case, DM is denser compared to the fiducial case. However, unlike the fermion case, the transition is now advanced. The net effect results in a larger $\gamma$ factor and a lower mass bound. At a high BEC fraction, the WDM mass bound can be arbitrarily low. This clearly shows that DM constituted of condensation, such as that of the fuzzy DM \citep{Ji:1994xh,Hu-Barkana-Gruzinov-PhysRevLett2000}, can have a mass that is orders of magnitude lower than the WDM mass bound for the standard thermal scenario. However, we can see from the lower panel of Figure \ref{fig:mass-constraints} that a high initial BEC fraction ($r_{\rm i}\gtrsim0.9$) is needed to lower the mass bound significantly. 

Eventually, at the low-mass region for the boson case with a very high or even a 100\% BEC component, the qWDM mass will be restricted due to the ``quantum pressure,'' such as the case for fuzzy DM \citep{Chavanis:2012aap,Irsic:2017yje,Dalal:2022rmp}, although the exact mass constraint is model-dependent \citep{Amin:2022nlh,Liu:2024pjg}. The study of this region is beyond the scope of this work.

% \subsection{A mapping of the mass constraint between qWDM and the thermal-relic WDM cases}
As shown in Figure \ref{fig:transfer-function}, the primary effect of quantum statistics on the transfer function is a change in the suppression wavenumber and consequently in $\ell_{\rm s}$. By aligning the transfer functions, we can establish a correspondence between the qWDM mass and the thermal relic WDM mass, which reads
\begin{equation}\label{eq:mass-mapping}
    m = 0.84\,m_{\textsc{wdm}}\left(\frac{2/g}{\gamma(\Delta_{\rm i},r_{\rm i})}\right)^{\nicefrac{1}{4}}\,. 
\end{equation}
The factor of $0.84$ arises from the fact that the suppression scale in the fiducial case with $m=0.84\,m_{\textsc{wdm}}$ matches the suppression scale in the thermal WDM case with $m_{\textsc{wdm}}$.

\section{Remarks on quantum statistical effects during the transition}
The degree of quantum statistical effects drops during RNRT, which can be readily seen with the conservation of specific entropy. When $\Delta_{\rm i}<0$, both the fermion and boson cases share similar behavior and reduce to the classical case in the $\Delta_{\rm i}\ll-1$ limit \citep{Lin:2023fao}, so we discuss here the situations where $\Delta_{\rm i}>0$ for the fermion case and $r_{\rm i}>0$ (with $\Delta_{\rm i}=0$) for the boson case.

For the fermion case, $\Delta$ drops from one constant $\Delta_{\rm i}$ in the relativistic phase to another constant $\Delta_{\rm f}$ in the nonrelativistic phase. Thus, the degenerate pressure can only appear today if it appeared in the early Universe. The drop of $\Delta$ is a result of the conservation of specific entropy. In both the relativistic and nonrelativistic phases, the specific entropy $s/n$ is a monotonic decreasing function only of $\Delta$. On the other hand, for a given $\Delta$, $s/n$ is smaller in the nonrelativistic phase than in the relativistic phase. Thus, to keep the specific entropy constant during RNRT, $\Delta$ must decrease. By equaling the relativistic and relativistic limits of $s/n$ [see Eq.\,\eqref{eq:specific-entropy-rel-nonrel-limits}], it can be shown that in the highly degenerate limit (i.e., $\Delta\rightarrow\infty$), $\Delta$ drops to half of its initial value, i.e., $\Delta_{\rm f}=\frac{1}{2}\Delta_{\rm i}$. So, if qWDM was initially in a highly degenerate state, then it remains in a highly degenerate state today.

For the boson case with an initial BEC component, i.e., $\Delta_{\rm i}=0$ and $r_{\rm i}>0$, the BEC fraction drops from one constant at the relativistic phase to another constant at the nonrelativistic phase. This is also a result of the conservation of specific entropy. Recall that it is $s/n_{\rm tot}=(1-r)s/n$ that remains a constant. We know that when $\Delta=0$, $s/n$ decreases during the transition. Thus, $r$ needs to decrease to keep $s/n_{\rm tot}$ constant. Therefore, qWDM can have a BEC component today only if it started with some BEC component. This agrees with the expectation in \citet{Fukuyama:2006gp}. Quantitatively, by equaling the initial and final specific entropy, one can show that the final BEC fraction $r_{\rm f}$ is related to the initial $r_{\rm i}$ by 
\begin{equation}\label{eq:final-BEC-fraction}
    r_{\rm f} = 1-\frac{1-r_{\rm i}}{1-r_{\rm i}^{\rm cr}}\,,
\end{equation}
with
\begin{equation}\label{eq:critical_BEC_fraction}
    r_{\rm i}^{\rm cr}=1-\frac{5J_{\nicefrac{5}{2}}^+(0)J_3^+(0)}{4J_{\nicefrac{3}{2}}^+(0)J_4^+(0)}\simeq0.6435\,
\end{equation}
being a critical initial BEC fraction. We have $r_{\rm f}>0$ if $r_{\rm i}>r_{\rm i}^{\rm cr}$. Thus, \textit{only when the initial BEC fraction is higher than $\sim$\,64\% will a BEC component remain today}. This conclusion holds even if the boson DM has some internal energy states. This is because Eq.\,\eqref{eq:final-BEC-fraction} is obtained based on the conservation of particle number and specific entropy, and the condensed component in some internal states does not contribute to the system's entropy. 

When $r_{\rm i}<r_{\rm i}^{\rm cr}$, the BEC fraction drops to $0$ at some point during RNRT, and then $\Delta$ drops from $0$ to a negative value until the nonrelativistic phase. On the other hand, if $r_{\rm i}\rightarrow1$, we also have $r_{\rm f}\rightarrow1$ from Eq.\,\eqref{eq:final-BEC-fraction}. Therefore, if qWDM started with a highly condensed state, it remains in a highly condensed state today.

\section{Summary and Conclusion}
In this work, we have studied the quantum statistical effects of a single-species WDM on the early-time background evolution of the density and pressure, cosmic structure growth, and the mass constraint from the Lyman-$\alpha$ forest. We have considered situations where the system is degenerate for a fermion DM or where a BEC component appears for a boson DM. We have paid special attention to the change of quantum statistical effects during RNRT, which affects the WDM sound horizon and the suppression scale of the matter power spectrum.

When quantum statistical effects are considered, two factors predominantly affect the constraint on the WDM mass from the cosmic structure. One is the qWDM number density compared to the thermal relic WDM case. The other is the time of the qWDM RNRT. Situations with a higher qWDM number density and an earlier transition make the mass lower bound weaker (i.e., the bound is lower). In both the degenerate fermion case and the boson case with a BEC component, the number density is higher than in the thermal relic WDM case. For the fermion case, the degenerate pressure postpones RNRT. The net effect is that the DM mass constraint is only slightly lower in the degenerate fermion case compared to that in the thermal relic WDM case. On the other hand, the boson case with a BEC component advances RNRT. As a result, the DM mass lower bound can be significantly relaxed. However, a large initial BEC fraction ($n_{\textsc{bec}}/n_{\rm tot}\gtrsim90\%$) is required to substantially weaken the mass lower bound. 

The degree of degeneracy for the fermion case and the BEC fraction for the boson case decrease during RNRT, both of which are results of the conservation of specific entropy.  In particular, for the boson case, the BEC fraction drops to zero unless the initial BEC fraction is higher than about 64\%. Small-scale problems have motivated proposals that involve boson DM with a completely or partially condensed state, but dynamically generating a BEC component does not achieve such a high initial fraction \citep{Madsen:1992am}. This does not constitute immediate trouble in scenarios where the boson DM somehow directly starts with a completely condensed state, as is what is taken for, e.g., axion DM \citep{Preskill:1982cy,Abbott:1982af,Dine:1982ah}, fuzzy DM \citep{Ji:1994xh,Hu-Barkana-Gruzinov-PhysRevLett2000} and others \citep{Fukuyama:2006gp}. However, it poses a question on how to generate a high DM BEC fraction dynamically. One way to avoid the above constraint is that the BEC state is generated during the nonrelativistic phase of DM. Another way is that small-scale BEC can be somehow formed during gravitational collapse. 

A caveat in this work is that we assume that qWDM is an ideal gas in an equilibrium phase-space distribution, so that the background evolution can be established in a parameterized way. This implicitly puts a constraint on the type and strength of the self-interaction. 
Considering more general self-interaction may result in different evolution \citep{Harko:2011zt,Guth:2014hsa}. Importantly, our main conclusions -- specifically, the variation in mass constraints with the degree of quantum statistics and the evolution of quantum statistics during RNRT -- remain qualitatively robust, provided that qWDM remains in equilibrium until the completion of RNRT. Moreover, the necessary self-interaction strength is significantly below the upper limits set by current astrophysical constraints; see Appendix \ref{sec:self-interaction}.

\begin{acknowledgements}
    W. L. thanks Vid Ir\v{s}i\v{c} for answering the questions about the unit of wavenumber used in the Lyman-$\alpha$ forest flux power spectrum, and Xingang Chen and Amol Upadh for useful feedback. W. L. acknowledges that this work is supported by the ``Science \& Technology Champion Project" (202005AB160002) and the ``Top Team Project" (202305AT350002), both funded by the ``Yunnan Revitalization Talent Support Program." This work is also supported by the ``Yunnan General Grant'' (202401AT070489).
\end{acknowledgements}

\appendix

\section{Methods}\label{sec:method-rel-nonrel-limits}
\subsection{Thermodynamical variables and their relativistic and their nonrelativistic limits}
To obtain the quantum statistical effects, we first calculate the background evolution of the qWDM density and pressure, then derive the adiabatic sound speed and implement it into the cosmic linear perturbation. Here, we highlight the key steps of the procedure used in \citet{Lin:2023fao}. With the momentum distribution given in Eq.\,\eqref{eq:HiDM-distribution}, the qWDM number density, energy density, and pressure can be calculated with
{
\allowdisplaybreaks
\begin{align}
n&= \int_0^\infty f(p) dp \,,\\
\rho &= \int_0^\infty f(p)\sqrt{p^2+m^2} dp\,, \\
P&=\int_0^\infty \frac{f(p)p^2}{3\sqrt{p^2+m^2}} dp\,.
\end{align}
}
By defining the mass-to-temperature ratio $x\equiv m/T$ and the effective chemical potential-to-temperature ratio $\Delta\equiv(\mu-m)/T$, the above can be cast into following forms:
\begin{align}
n &= \frac{n_*}{x^3}\mathcal{N}\,~~\textrm{with~~~}\mathcal{N}(x,\Delta)=\int_0^\infty\frac{dzz^2}{\exp(\sqrt{z^2+x^2}-x-\Delta)\pm1}\,, \label{eq:HiDM-number-density}\\
\rho &= \frac{n_*m}{x^4}\mathcal{R}\,~~\textrm{with~~~}\mathcal{R}(x,\Delta)=\int_0^\infty\frac{dzz^2\sqrt{z^2+x^2}}{\exp(\sqrt{z^2+x^2}-x-\Delta)\pm1}\,\label{eq:HiDM-energy-density}\\
P&=\frac{n_*m}{x^4}\mathcal{P}\,~~\textrm{with~~~}\mathcal{P}(x,\Delta)=\frac{1}{3}\int_0^\infty\frac{dzz^4/\sqrt{z^2+x^2}}{\exp(\sqrt{z^2+x^2}-x-\Delta)\pm1}\,,\label{eq:HiDM-pressure}
\end{align}
where $n_*=\frac{gm^3}{2\pi^2\hbar^3}$. The relativistic and nonrelativistic limits of Eqs.\,\eqref{eq:HiDM-number-density}-\eqref{eq:HiDM-pressure} can be calculated analytically, which we give in Table \ref{tab:rel_and_nonrel_limits}. Note that the condition for the nonrelativistic limit is $x\gg1$ and $x\gg\Delta$ as we explained.

\begin{table}[]
\begin{ruledtabular}
\renewcommand{\arraystretch}{2.}
    \begin{tabular}{c|l|l}
     & The relativistic limit, $x\ll1$ & The nonrelativistic limit, $x\gg1$ and $x\gg\Delta$ \\
     \hline
    & $\mathcal{N}=J_3^\mp(\Delta)+2xJ_2^\mp(\Delta) + \mathcal{O}(x^2)$ & $\mathcal{N}=\frac{1}{2}(2x)^{\nicefrac{3}{2}}\left(J_{\nicefrac{3}{2}}^\mp(\Delta)+\frac{5}{4x}J_{\nicefrac{5}{2}}^\mp(\Delta)+\mathcal{O}(\frac{1}{x^2})\right)$ \\
    & $\mathcal{R}=J_4^\mp(\Delta)+3xJ_3^\mp(\Delta)+\mathcal{O}(x^2)$ & $\mathcal{R}=\frac{1}{2}x\cdot(2x)^{\nicefrac{3}{2}}\left(J_{\nicefrac{3}{2}}^\mp(\Delta)+\frac{9}{4x}J_{\nicefrac{5}{2}}^\mp(\Delta)+\mathcal{O}(\frac{1}{x^2})\right)$   \\
    &  $\mathcal{P}=\frac{1}{3}\mathcal{R}+\mathcal{O}(x^2)$ &  $\mathcal{P}=\frac{1}{3}(2x)^{\nicefrac{3}{2}}\left(J_{\nicefrac{5}{2}}^\mp(\Delta)+\mathcal{O}(\frac{1}{x})\right)$ \\
    & $s/n=\frac{4J_4^\mp(\Delta)}{3J_3^\mp(\Delta)}-\Delta+\mathcal{O}(x^2)$  &  $s/n=\frac{5J_{\nicefrac{5}{2}}^\mp(\Delta)}{3J_{\nicefrac{3}{2}}^\mp(\Delta)}-\Delta+\mathcal{O}(\frac{1}{x^2})$ 
         \\ \hline
    $\Delta\leq0$ & \multicolumn{2}{c}{$J_s^\mp(\Delta)=\Gamma(s)f_s^\mp(\Delta)\exp(\Delta)$}\\
    %$J_s^\mp(\Delta)=\Gamma(s)f_s^\mp(\Delta)\exp(\Delta)$ & $I_s^\mp(\Delta)=\frac{1}{2}\Gamma(s+\frac{1}{2})f_{s{\tiny+\nicefrac{1}{2}}}^\mp(\Delta)\exp(\Delta)$ \\
    %& $J_3^{\mp}(\Delta)=\Gamma(3)f_3^\mp(\Delta)\exp(\Delta)$ & $J_{\nicefrac{3}{2}}^\mp(\Delta) = \frac{1}{2}\Gamma(3/2)f_{\nicefrac{3}{2}}^\mp(\Delta)\exp(\Delta)$ \\ 
     %& $J_4^{\mp}(\Delta)=\Gamma(4)f_4^\mp(\Delta)\exp(\Delta)$ & $J_{\nicefrac{5}{2}}^\mp(\Delta) = \frac{1}{2}\Gamma(5/2)f_{\nicefrac{5}{2}}^\mp(\Delta)\exp(\Delta)$ \\
     \hline
    $\Delta>0$   & %$J_2^-(\Delta)=\frac{1}{2}\Delta^2+\frac{\pi^2}{6}-f_2^-(-\Delta)\exp(-\Delta)$  
    $J_s^-(\Delta) = \frac{1}{s}\Delta^s+\sum\limits_{k=1}^{k\leq \frac{s}{2}} A_s^kJ_{2k}^-(0)\Delta^{s-2k}+(-1)^{s-1}J_s^-(-\Delta)$
    &  $\displaystyle J_{\nicefrac{3}{2}}^-(\Delta) = \frac{2}{3}\Delta^{\nicefrac{3}{2}}+\frac{\sqrt{\pi}}{2}\sum_{n=1}^{\infty}\frac{(-1)^{n-1}}{n^{\nicefrac{3}{2}}}\times$~~~~~~~  \vspace{-5pt}\\ fermion & &\hfill  $\Big( {\rm erfi}(\sqrt{n\Delta})e^{-n\Delta}  +{\rm erfc}(\sqrt{n\Delta})e^{n\Delta}\Big)$  \\
      & ~~~~~~~~~where~~~ $A_s^k=\frac{2\times(s-1)!}{(2k-1)!(s-2k)!}$
      %$J_3^-(\Delta)=\frac{1}{3}\Delta^3+\frac{\pi^2}{3}\Delta+2f_3^-(-\Delta)\exp(-\Delta)$  
      & $\displaystyle J_{\nicefrac{5}{2}}^-(\Delta) = \frac{2}{5}\Delta^{\nicefrac{5}{2}}+\frac{\pi^2}{4}\Delta^{\nicefrac{1}{2}}+\frac{3\sqrt{\pi}}{4}\sum_{n=1}^{\infty}\frac{(-1)^n}{n^{\nicefrac{5}{2}}} \times$~~~~~~~~~~~~ \vspace{-5pt}\\
    & &    $\hfill\Big( {\rm erfi}(\sqrt{n\Delta})e^{-n\Delta}-{\rm erfc}(\sqrt{n\Delta})e^{n\Delta}\Big)$ \\
    % & $J_4^-(\Delta)=\frac{1}{4}\Delta^4+\frac{\pi^2}{2}\Delta^2+\frac{7\pi^4}{60}-6f_4^-(-\Delta)\exp(-\Delta)$  & 
    \end{tabular}
    \end{ruledtabular}
    \caption{Summary of the relativistic and the nonrelativistic limits of the characteristic functions of thermal number density ($n = \frac{gm^3}{2\pi^2\hbar^3}\frac{1}{x^3}\mathcal{N}$), energy density ($\rho= \frac{gm^4}{2\pi^2\hbar^3}\frac{1}{x^4}\mathcal{R}$), pressure ($p=\frac{gm^4}{2\pi^2\hbar^3}\frac{1}{x^4}\mathcal{P}$), and specific entropy $s/n$ ($n$ is the thermal number density only).  The relativistic (nonrelativistic) limit is defined as $x\ll1$ ($x\gg1$ and $x\gg\Delta$). 
    The function $f^{\mp}_s$ is defined as $f^{\mp}_s(\Delta)\equiv\sum_{n=0}^\infty\frac{(\mp1)^{n}}{(n+1)^s}\exp(n\Delta)=\Phi(e^{\mp\Delta},s,1)$, where now $-$ is for fermion, $+$ for boson, and $\Phi$ is the Lerch transcendent. The functions ${\rm erfi}$ and $\rm{erfc}$ are the complementary and imaginary error functions. For the boson case with a BEC component, the above equations apply to the thermal component with $\Delta=0$. The integration forms of the $J$'s functions are defined as $J_s^\mp(\Delta)\equiv\int_0^\infty\frac{z^{s-1}{\rm d}z}{\exp(z-\Delta)\pm1}$. Some important properties are $\frac{{\rm d}}{{\rm d}\Delta}J_s^\mp(\Delta)=(s-1)J_{s-1}^\mp(\Delta)$. Note that the subscript $s$ is an integer or half-integer not to be confused with the entropy density.}
    \label{tab:rel_and_nonrel_limits}
\end{table}

\begin{figure}
    \centering
    \includegraphics[width=0.9\linewidth]{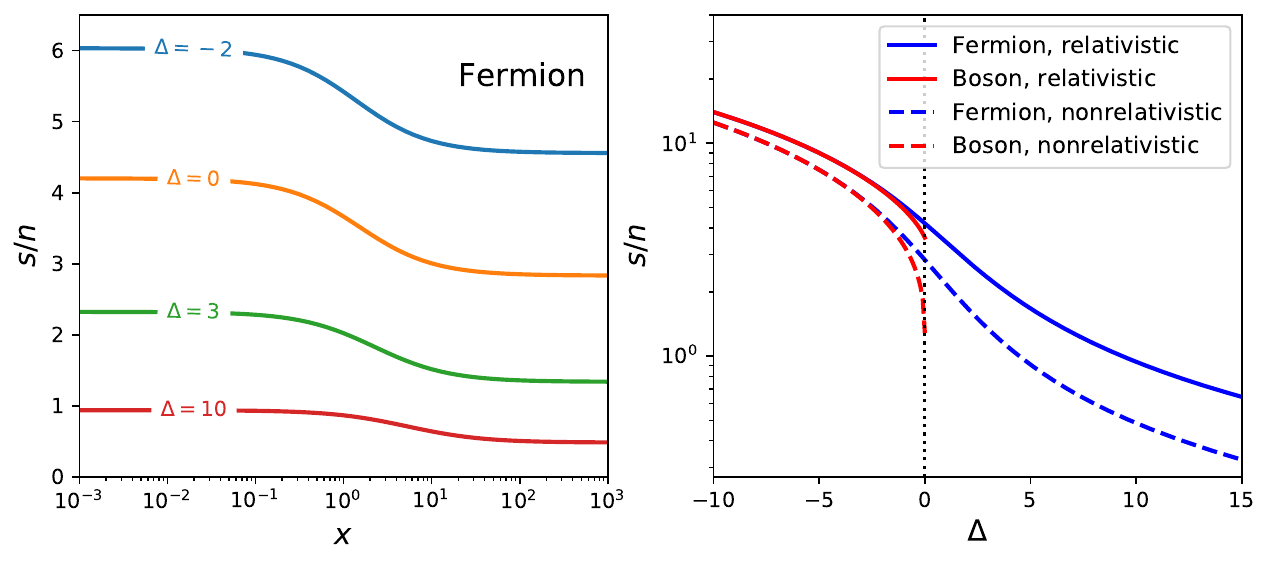}
    \caption{The dependence of thermal specific entropy $s/n$ on the mass-to-temperature ratio $x$ and the effective chemical potential-to-temperature ratio $\Delta$. Left: curves of $s/n$ as a function of $x$ for some fixed $\Delta$s for the fermion case. For a given $\Delta$, $s/n$ drops from a constant in the relativistic limit to another constant in the nonrelativistic limit. Right: curves of $s/n$ as a function of $\Delta$ in the relativistic limit (solid) and in the nonrelativistic limit (dashed). The blue curves are for fermion and the red curves are for boson. Note that, for the boson case, the specific entropy concerning the total particle number is $s/n_{\rm tot}=(1-r)s/n.$}
    \label{fig:specific-entropy}
\end{figure}

From Euler's theorem on homogeneous functions, the entropy of an ideal gas is given by $TS=(U+VP-\mu N)$ for given volume $V$, energy $U$ (including the rest energy), and particle number $N$. Dividing it by $V$ and $T$, we obtain the entropy density, which reads
\begin{equation}\label{eq:HiDM-entropy-density}
    s= \frac{\rho+P-\mu n}{T}\,.
\end{equation}
Then, with Eqs.\,\eqref{eq:HiDM-number-density}-\eqref{eq:HiDM-pressure} the specific entropy is then given by
\begin{equation}\label{eq:HiDM-specific-entropy}
    \frac{s}{n}=\frac{\mathcal{R}(x,\Delta)+\mathcal{P}(x,\Delta)}{\mathcal{N}(x,\Delta)}-x-\Delta\,.
\end{equation}
Note that while Eq.\,\eqref{eq:HiDM-specific-entropy} still holds for the boson case with a BEC component, with $n$ being the density of the thermal component, the correct specific entropy that conserves is $s/n_{\rm tot}=(1-r)s/n$. We show the dependence of $s/n$ on $x$ and $\Delta$ in Figure \ref{fig:specific-entropy}. For a given $\Delta$, the value of $n/s$ is a constant in the relativistic limit and is a smaller constant in the nonrelativistic limit. The large $|\Delta|$ behaviors of the relativistic and nonrelativistic limits of $s/n$ are useful and given by (positive $\Delta$ only applies to the fermion case),
\begin{equation}\label{eq:specific-entropy-rel-nonrel-limits}
s/n = 
\begin{cases}
    \frac{4J_4^\mp(\Delta)}{3J_3^\mp(\Delta)}-\Delta & \mathrlap{(\text{for } x\ll1)}\hphantom{(\text{for } x\gg1~\&~x\gg\Delta)}~~=
    \begin{cases}
        \mathrlap{\frac{\pi^2}{\Delta}}\hphantom{-\Delta+5/2} & \text{for } x\ll1~\&~\Delta\gg1 \text{, relativistic and degenerate,}\\
        -\Delta+4  &\text{for } x\ll1~\&~\Delta\ll-1 \text{, relativistic and classical,}
    \end{cases}   \vspace{12pt}  \\ 
    \frac{5J_{\nicefrac{5}{2}}^\mp(\Delta)}{3J_{\nicefrac{3}{2}}^\mp(\Delta)}-\Delta & (\text{for } x\gg1~\&~x\gg\Delta)~~=
    \begin{cases}
        \frac{\pi^2}{2\Delta} &\text{for } x\gg\Delta\gg1  \text{, nonrelativistic and degenerate,}\\
        -\Delta+5/2 &\text{for } x\gg1~\&~\Delta\ll-1\text{, nonrelativistic and classical.}
    \end{cases}
\end{cases}
\end{equation}

\subsection{The background evolution}
The two independent dynamical variables are $x$ and $\Delta$ ($x$ and $r$ for the boson case with $\Delta=0$), whose background evolution is derived using the conservation of the comoving particle number and the conservation of the specific entropy (concerning the total particle number), that is, 
\begin{align}
    {\rm  d} (n_{\rm tot}a^3) = 0\,,\label{eq:conservation_number_diff}\\
    {\rm d} (s/n_{\rm tot}) =0\,.\label{eq:conservation_specific_entropy_diff}
\end{align}
The above conditions set up the differential equations for $x$ and $\Delta$ as functions of the scale factor $a$, which, for the fermion case and the boson case without condensation, read
\begin{equation}\label{eq:differential-x-Delta}
    \begin{pmatrix}{\rm d}\ln x\\ {\rm  d}\Delta \end{pmatrix} = \begin{pmatrix}
    1-\frac{x}{3\mathcal{N}}\mathcal{N}_x  &  -\frac{1}{3\mathcal{N}}\mathcal{N}_\Delta \\
    \frac{x}{\mathcal{N}}\xi_x-\frac{x\,\xi}{\mathcal{N}^2}\mathcal{N}_x-x & ~~\frac{1}{\mathcal{N}}\xi_\Delta-\frac{\xi}{\mathcal{N}^2}\mathcal{N}_\Delta-1
    \end{pmatrix}^{-1} \begin{pmatrix}{\rm  d}\ln a\\ 0 \end{pmatrix}\,,
\end{equation}
where $\xi=\mathcal{R}+\mathcal{P}$ and the subscripts denote partial derivatives (e.g., $\xi_x\equiv\frac{\partial\xi}{\partial x}$). For the boson case with a BEC component, qWDM begins with $\Delta=0$ and $r>0$. Then, Eqs.\,\eqref{eq:conservation_number_diff} and \eqref{eq:conservation_specific_entropy_diff} reduce to a differential equation for $x$,
\begin{equation}\label{eq:differential-x-BEC}
    \frac{{\rm  d}\ln x}{{\rm  d}\ln a} =  \frac{3(x\mathcal{N}-\xi)}{2x\mathcal{N}-x^2\mathcal{N}_x-3\xi+x\xi_x}\bigg|_{\Delta=0}\,.
\end{equation}
We integrate the above equation to find $x(a)$, which is then substituted into $(1-r)s/n={\rm constant}$ to give $r(a)$. If the initial BEC fraction is $r_{\rm i}>r_{\rm cr}$ [see Eq.\,\eqref{eq:critical_BEC_fraction}], the integration of Eq.\,\eqref{eq:differential-x-BEC} proceeds until $r=0$, and we continue to solve Eq.\,\eqref{eq:differential-x-Delta} with the initial value of $x$ given by that at $r=0$. Otherwise, the integration of Eq.\,\eqref{eq:differential-x-BEC} proceeds until $a=1$.

The initial conditions are described as follows. For both the fermion and boson cases, the value of $x/a$ is a constant in the early Universe. Thus, we specify the value of $x/a$ in the relativistic limit as the first initial condition. We call $T_0/m\equiv\lim_{a\rightarrow 0} a/x$ the initial ``warmness'' of qWDM. The second initial condition is treated differently for two possibilities. For the fermion case and the boson case without a BEC component, the value of $\Delta$ is a constant in the early Universe. In this case, we specify $\Delta$ in the relativistic limit ($\Delta_{\rm i}$) as the second initial condition. For the boson case with $\Delta=0$, a BEC component appears. The BEC fraction $r$ is constant in the early Universe. In that case, we use $r$ in the relativistic limit ($r_{\rm i}$) as the second initial condition. That is, the initial conditions are specified by the set of $(x_{\rm i}/a_{\rm i},\,\Delta_{\rm i})$, except for the boson case with $\Delta_{\rm i}=0$ where they are specified by $(x_{\rm i}/a_{\rm i},\,r_{\rm i})$.

With the background evolution of $x$ and $\Delta$ obtained above, we substitute them into Eqs.,\eqref{eq:HiDM-number-density}-\eqref{eq:HiDM-pressure} to get the background evolution of the number density, energy density, and pressure of qWDM. We assume that the qWDM perturbation is an adiabatic process with the adiabatic sound speed calculated by Eq.,\eqref{eq:adiabatic-sound-speed}. With a finite equation of state and a finite adiabatic sound speed, the qWDM linear perturbation equations are modified compared to the cold DM case and read \citep{Ma-Bertschinger1995},
\begin{align}
    \delta_{{\rm q}{\textsc{wdm}}}'&=-kq_{{\rm q}{\textsc{wdm}}}-(1+w_{{\rm q}{\textsc{wdm}}})h_{\rm{s}}'/2+3\mathcal{H}(w_{{\rm q}{\textsc{wdm}}}-c_{\rm s}^2)\delta_{{\rm q}{\textsc{wdm}}}\,,\label{eq:HiDM-delta-dot}\\
    q_{{\rm q}{\textsc{wdm}}}'&=(3w_{{\rm q}{\textsc{wdm}}}-1)\mathcal{H}q_{{\rm q}{\textsc{wdm}}}+kc_{\rm s}^2\delta_{{\rm q}{\textsc{wdm}}}\,,\label{eq:HiDM-velocity-dot}
\end{align}
where $\delta_{{\rm q}{\textsc{wdm}}}$ and $q_{{\rm q}{\textsc{wdm}}}$ are the qWDM overdensity and heat flux, $'$ denotes the derivative with respect to the conformal time, $w_{{\rm q}{\textsc{wdm}}}\equiv\frac{P}{\rho_{\rm tot}}$, $\mathcal{H}$ is the conformal Hubble parameter, and $h_{\rm s}$ and $\eta_{\rm s}$ are the two synchronous gauge gravitational variables \citep{Ma-Bertschinger1995}. We have omitted the pressure anisotropy in Eq.,\eqref{eq:HiDM-velocity-dot} since we assume that qWDM has an equilibrium phase-space distribution that is isotropic. We implement the above linear perturbation into the public code \textsc{camb} \citep{Lewis-Challinor-Lasenby-2000} to calculate the linear matter power spectrum and determine the suppression scale using Eq.\,\eqref{eq:suppression-scale-camb}.

\section{The self-interaction strength}\label{sec:self-interaction}
We have investigated the effects of quantum statistics, including the shift of time of RNRT and the corresponding impacts in the matter power spectrum. For those effects to hold, a certain level of self-interaction is required to maintain DM equilibrium until RNRT is fully completed. Specifically, DM must remain in equilibrium until a scale factor $a_{\rm f}$ sufficiently later than the time of RNRT, such as $a_{\rm f}>10a_{\rm nr}$. The necessary self-interaction strength depends on the type of interaction. In this context, we consider a DM self-interaction with a constant cross section $\sigma_{\textsc{si}}$ to demonstrate that this requirement can be met without violating current astrophysical and cosmological constraints. 

\begin{figure}
    \centering
    \includegraphics[width=0.95\linewidth]{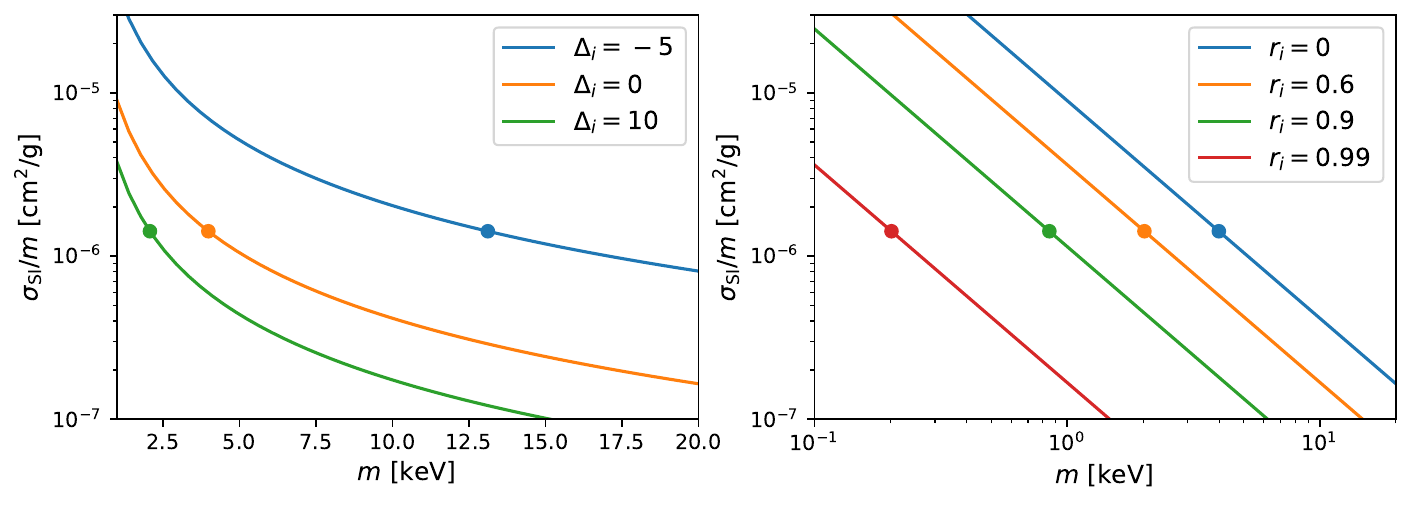}
    \caption{The minimum constant cross section required for $a_{\rm f}>10a_{\rm nr}$ for the fermion case (left) and the boson case with BEC (right). Different curves correspond to different degrees of degeneracy for the fermion cases and different BEC fraction for the boson cases. The dots on curves mark the corresponding mass low-bound from the Lyman-$\alpha$ forest for different cases. The required minimum cross section is well below the astrophysical upper bounds ranging $0.2\sim0.5$ cm$^2$/g \citep{Randall:2008ppe,Eckert:2022qia}. Considering a velocity-dependent cross section with $\sigma_{\textsc{si}}\propto \langle v\rangle^{-p}$ for some positive $p$ raises the astrophysical constraints on the cross section in the near-relativistic regime.} 
    \label{fig:interaction}
\end{figure}

We estimate $a_{\rm f}$ by equaling the DM self-interaction rate per particle in thermal states with the Hubble rate, i.e., $\sigma_{\textsc{si}}\langle v\rangle n_{\rm tot}\big|_{a_{\rm f}} = H\big|_{a_{\rm f}}$. Like the sound speed, we parameterized average velocity at the nonrelativistic phase as $\langle v\rangle=\frac{a_{\rm nr}}{\sqrt{3}a}$. With $a_{\rm nr}$ solved in Eq.\,\eqref{eq:DM-fraction-a-hat-cs}, it can be shown that 
\begin{equation}\label{eq:a_f}
    a_{\rm f} = 5.6\times10^{-4}\left(\frac{2/g}{\gamma(\Delta_{\rm i},r_{\rm i})}\right)^{\nicefrac{1}{6}}\left(\frac{\Omega_{\rm dm}h^2}{0.12}\right)^{\nicefrac{2}{3}}\left(\frac{\sigma_{\textsc{si}}/m}{1\,{\rm{cm^2/g}} }\right)^{\nicefrac{1}{2}}\left(\frac{1\,\rm{keV}}{m}\right)^{\nicefrac{2}{3}}\,.
\end{equation}
The above result is consistent with that given in \citet{Garani:2022yzj} in the fiducial case where $\Delta_{\rm i}=0$ and $r_{\rm i}=0$ but is generalized to include scenarios involving degenerate fermions and bosons with BEC. The minimally required interaction strength is shown in Figure \ref{fig:interaction}, presented by $\sigma_{\textsc{si}}/m$ as a function of qWDM mass. The minimally required cross section varies with the degree of degeneracy for the fermion case and the BEC fraction for the boson case. Considering the mass constraints from the Lyman-$\alpha$ forest, the minimally required $\sigma_{\textsc{si}}/m$ for the lowest allowed mass is about $1.4\times10^{-6}~{\rm cm}^2/{\rm g}$ (see the height of the dots in Figure \ref{fig:interaction}) and is lower for a higher DM particle mass. These are well below the upper bound from the current astrophysical constraints $0.2\sim0.5~{\rm cm}^2/{\rm g}$ \citep{Randall:2008ppe,Eckert:2022qia}. 

In the above discussion, we have assumed a velocity-independent cross section. However, the conclusion that the required cross section well satisfies the current astrophysical constraints still holds as long as $\sigma_{\textsc{si}}\propto \langle v\rangle^{-p}$ with some positive $p$ for the following reason. Since we require DM remain in equilibrium until RNRT is complete, the above required cross section should be interpreted as that for relativistic or nearly relativistic case. The current astrophysical constraints are at $v\sim 1000$ km/s for the case in \citet{Randall:2008ppe} and at $v\sim 5000$ km/s for the case in \citet{Eckert:2022qia}. These will be translated into a higher upper bound of $\sigma_{\textsc{si}}/m$ in the nearly-relativistic case. Compared to current astrophysical constraints, only a weak self-interaction is necessary to maintain the equilibrium of qWDM until RNRT is complete. This is primarily due to the high number density of qWDM in the early Universe, which results in a significant self-interaction rate, even with a small cross section. Consequently, the strong self-interactions often assumed for WDM would keep it in its own thermal equilibrium, preventing it from behaving as a free-streaming particle.

%, we shall call $a/x$ in the relativistic limit the initial warmness. 

%Notes: \citet{2013-Destri-DeVega-Sanchez-DMHalo},
%\citet{Moore1994,Oh:2015xoa,2015-Chavanis-Lemou-Mehats} for DM halo

\bibliography{HDMreference_II}{}
\bibliographystyle{aasjournal}

%% This command is needed to show the entire author+affiliation list when
%% the collaboration and author truncation commands are used.  It has to
%% go at the end of the manuscript.
%\allauthors

%% Include this line if you are using the \added, \replaced, \deleted
%% commands to see a summary list of all changes at the end of the article.
%\listofchanges

\end{document}